\documentclass[onecolumn]{aa}
\usepackage{graphicx}
\usepackage{txfonts}
\begin{document}
   \title{Embedded disks in Fornax dwarf elliptical
galaxies \thanks{Based on observations collected at the European
Southern Observatory, Chile (ESO Large Programme Nr.~165.N-0115)}}

   \author{S. De Rijcke \inst{1}\fnmsep\thanks{Postdoctoral Fellow of the Fund for Scientific
Research - Flanders (Belgium)(F.W.O)},
	   H. Dejonghe \inst{1},
           W.~W. Zeilinger \inst{2}
	\and
	   G.~K.~T. Hau \inst{3}
          }

   \offprints{S. De Rijcke}

   \institute{Sterrenkundig Observatorium, Ghent University,
		Krijgslaan 281, S9 \\ \email{sven.derijcke@rug.ac.be},
		\\ \email{herwig.dejonghe@rug.ac.be} \and Institut
		f\"ur Astronomie, Universit\"at Wien,
		T\"urkenschanzstra{\ss}e 17, A-1180 Wien, Austria \\
		\email{zeilinger@astro.univie.ac.at} \and {\tt ESO},
		Alonso de Cordova 3107, Santiago, Chile \\
		\email{ghau@eso.org} }

   \date{Received ; accepted }

\abstract{We present photometric and kinematic evidence for the
presence of stellar disks, seen practically edge-on, in two Fornax
dwarf galaxies, FCC204 (dS0(6)) and FCC288 (dS0(7)). This is the first
time such structures have been identified in Fornax dwarfs. FCC2088
has only a small bulge and a bright flaring and slightly warped disk
that can be traced out to $\pm 23${\arcsec} from the center (2.05~kpc
for $H_0=75$~km/s/Mpc). FCC204's disk can be traced out to $\pm
20${\arcsec} (1.78~kpc). This galaxy possesses a large bulge. These
results can be compared to the findings of Jerjen {\em et al.}
(\cite{jer00}) and Barazza {\em et al.}  (\cite{bar02}) who discovered
nucleated dEs with spiral and bar features in the Virgo Cluster.
\keywords{dwarf galaxies -- Fornax Cluster -- individual
galaxies~:~FCC204, FCC288}}

\maketitle
%

\section{Introduction}


Dwarf ellipticals (dEs) seem to prefer the densest regions of the
universe and are found abundantly in galaxy clusters and groups
(e.g. Conselice {\em et al.}, \cite{con01}; Ferguson~\&~Sandage,
\cite{fer89b}; Binggeli {\em et al.}, \cite{bin87}). This circumstance
most likely has important consequences for their evolution. Moore {\em
et al.}  (\cite{moo98}) have shown how late-type disk galaxies that
orbit in a cluster can loose angular momentum by interactions with
massive galaxies and, to a lesser degree, by tidal forces induced by
the cluster potential. $N$-body simulations performed by Mayer {\em et
al.}  (\cite{may01}) show that small disk galaxies that are close
companions to a massive galaxy will be affected likewise. A small
disk galaxy is destabilized and develops a bar that gradually slows
down by dynamical friction, transporting angular momentum to the halo
and to stars at larger radii. Since the latter are being stripped,
angular momentum is lost. Gas is funneled in towards the center by
torques exerted by the bar where it is converted into stars, thus
forming a nucleus. The small companion is heated by the subsequent
buckling of the bar (e.g. Merrifield~\&~Kuijken (\cite{mer99}) and
references therein) and by bending modes of the disk and is
transformed from a rotationally-flattened object into an anisotropic,
slowly rotating spheroidal galaxy. The effect on a dwarf galaxy
depends on its orbit through a cluster or around a massive
companion. For instance, retrograde interactions have a much less
damaging effect than prograde ones and may even preserve some of the
initial disk structure. Thus, these simulations allow for the
existence of fast-rotating dwarfs and for dEs that still contain a
stellar disk.

Recently, fast-rotating dEs have been discovered by e.g. De Rijcke
{\em et al} (\cite{rij01}) and Simien~\&~Prugniel
(\cite{sim02}). Jerjen {\em et al.} (\cite{jer00}) and Barazza {\em et
al.}  (\cite{bar02}) discovered spiral structure and bars in 5 bright
nucleated dEs in the Virgo Cluster~:~IC3328, IC0783, IC3349, NGC4431,
and IC3468. Ryden {\em et al.} (\cite{ryd99}) also report dEs with
disky isophotes in the Virgo cluster. Interpreting small number
statistics, about 20\% of the Virgo dEs could harbor embedded spiral
or bar systems.

In this paper, we present photometric as well as kinematic evidence
for the presence of stellar disks in two Fornax dwarf galaxies, FCC204
(dS0(6)) and FCC288 (dS0(7)). In the next section, the photometry of
both objects is discussed. In section \ref{kin}, we present their
major-axis stellar kinematics. Our conclusions are summarized in
section \ref{con}.

\section{Surface Photometry} \label{phot}


\begin{figure*}
\vspace*{8cm}
\special{hscale=80 vscale=80 hsize=700 vsize=220
hoffset=-95 voffset=-70 angle=0 psfile="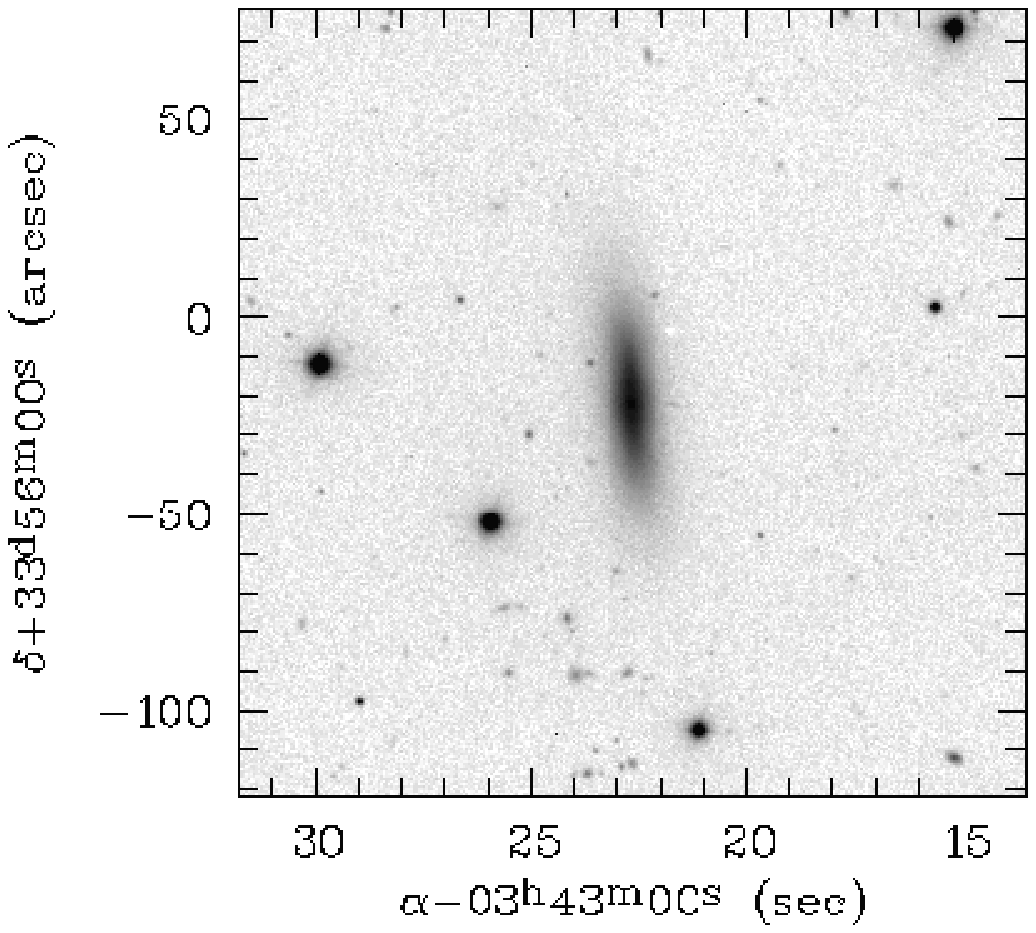"}
\special{hscale=82 vscale=82 hsize=700 vsize=220
hoffset=180 voffset=-72 angle=0 psfile="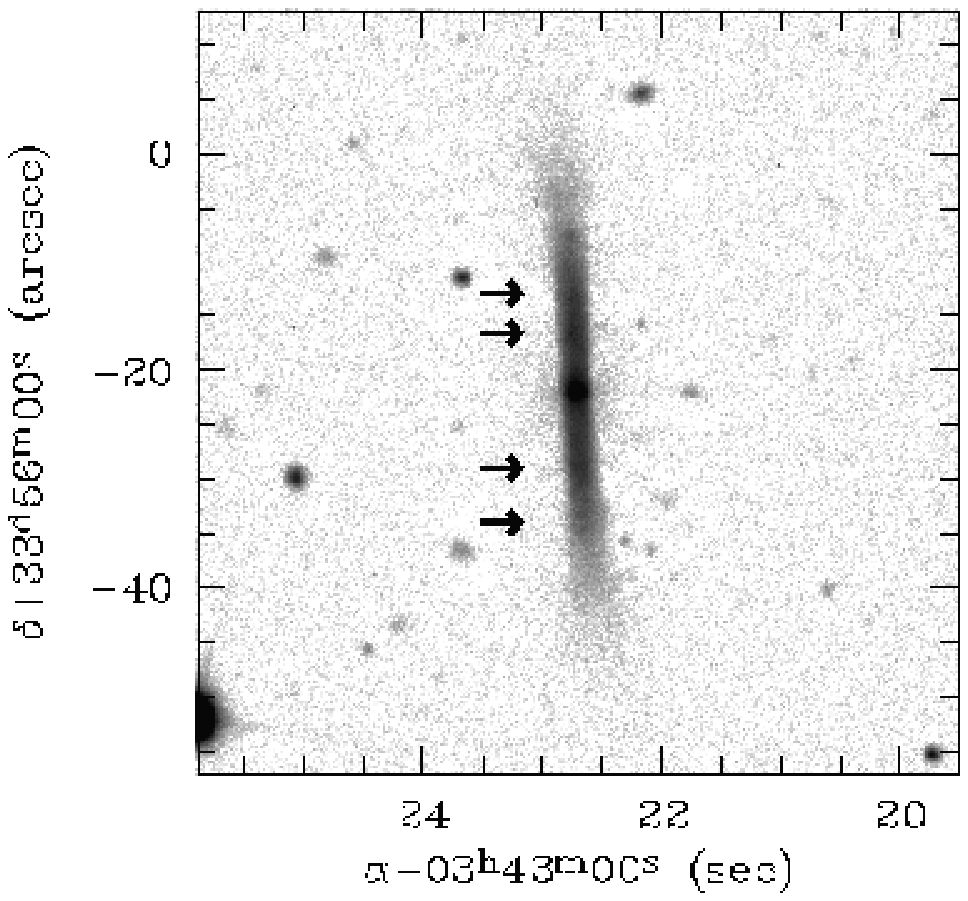"}
\caption{Left panel~:~R-band image of FCC288. Right panel~:~result of
unsharp masking. The disk embedded in FCC288 runs practically across
the whole face of the galaxy. The flaring of the disk and the
brightness fluctuations in it (marked by arrows) are clearly
visible. The disk is slightly warped with the north side tending
towards the east. The bulge is fairly small, reminiscent of those in
late-type spirals. \label{imafcc288}}
\end{figure*}

\begin{figure*}
\vspace*{8cm}
\special{hscale=80 vscale=80 hsize=700 vsize=220
hoffset=-95 voffset=-70 angle=0 psfile="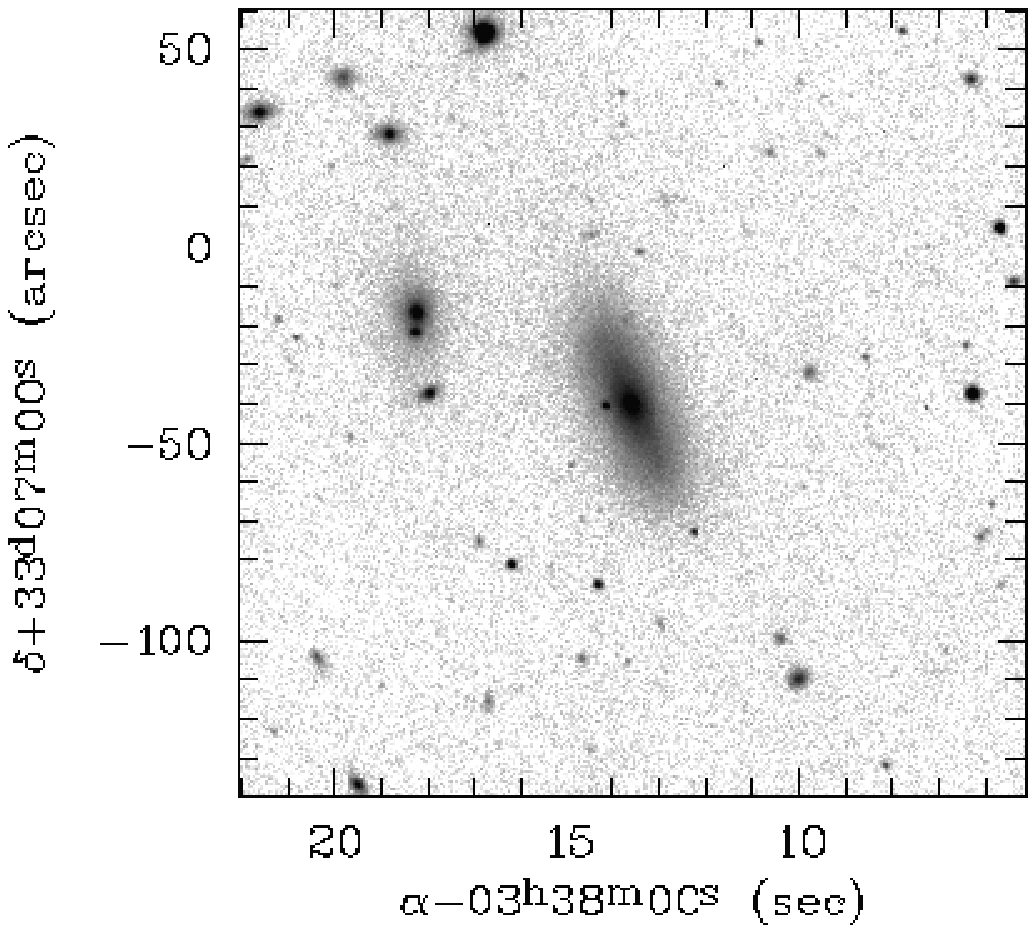"}
\special{hscale=82 vscale=82 hsize=700 vsize=220
hoffset=180 voffset=-72 angle=0 psfile="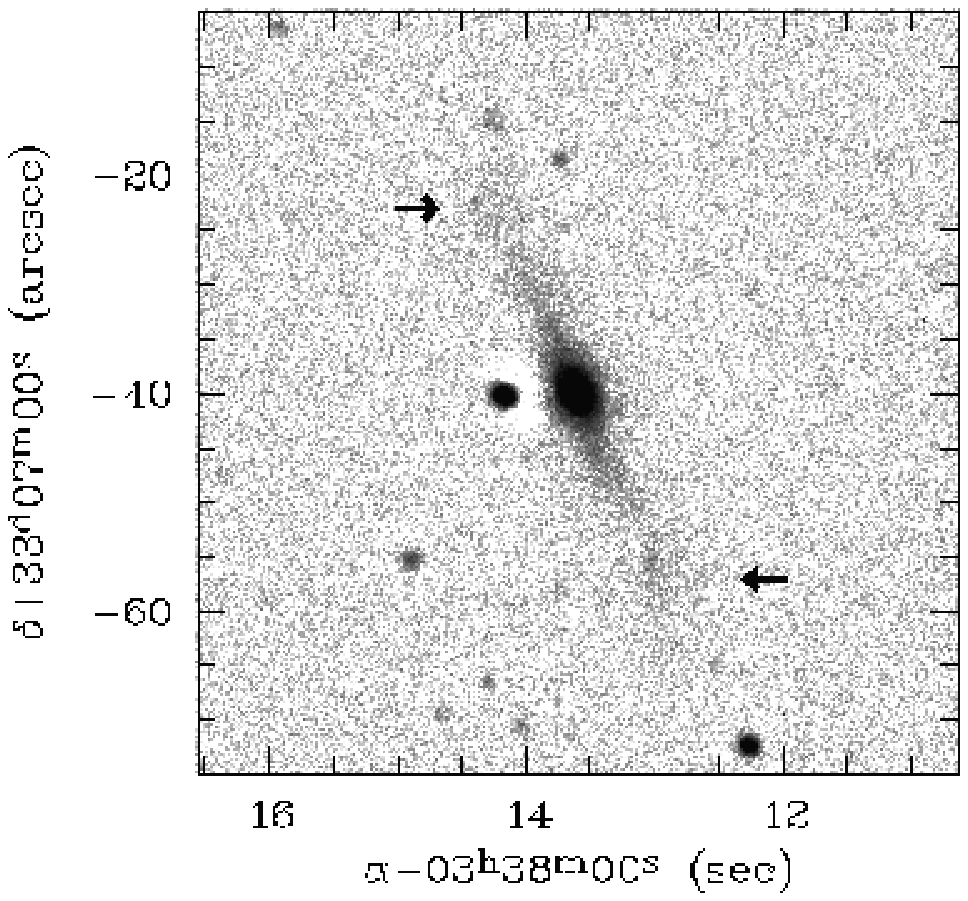"}
\caption{Left panel~:~R-band image of FCC204. Right panel~:~result of
unsharp masking. Though much less obvious than in FCC288, a disk can
still be discerned. This object has a more massive bulge than
FCC288. At the outer edges of the disk, two brightness peaks are
visible (marked by arrows). \label{imafcc204}}
\end{figure*}

\begin{figure}[ht]
\vspace*{7cm}
\special{hscale=45 vscale=45 hsize=700 vsize=200
hoffset=20 voffset=230 angle=-90 psfile="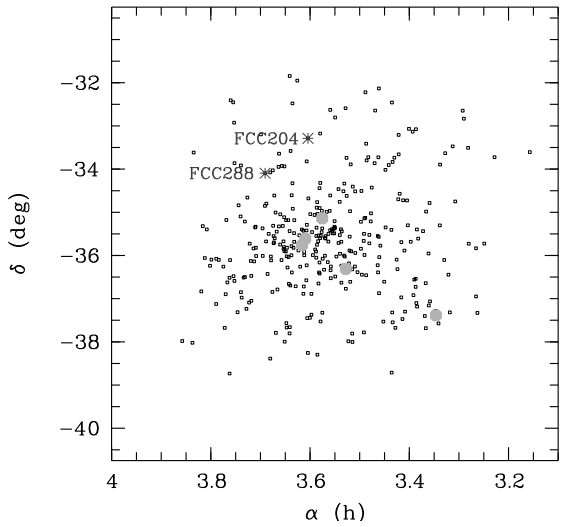"}
\caption{Position of FCC204 and FCC288 in the Fornax Cluster
(asterisks). The dots mark the positions of all 340 galaxies in the
Fornax Cluster Catalog. The brightest cluster members ($M_B<-20$) are
plotted as grey dots. \label{pos}}
\end{figure}

We collected R and I band images of a sample of Fornax dEs with {\tt
FORS2} on Kueyen ({\tt VLT-UT2}) in the period 1-8/11/2000. The CCD
pixel-scale is 0.2~{\arcsec}. A 600~sec R-band image of FCC288 and a
120~sec R-band image of FCC204 are presented in Figures
\ref{imafcc288} and \ref{imafcc204}, respectively. The average seeing
was $0.9${\arcsec}~FWHM (judged from a number of stars in the R band
images) for both galaxies. Basic photometric parameters can be found
in Table 1. In Figure \ref{pos}, the positions of FCC204 and FCC288
are plotted along with all other likely cluster members taken from the
Fornax Cluster Catalog (Ferguson \cite{fer89a}). Drinkwater {\em et
al.} (\cite{dri01}) spectroscopically confirmed FCC204 and FCC288 as
cluster members. These authors did not detect significant Halpha
emission, corroborating the classification of these galaxies as dS0s
and not as dwarf spirals. Both objects are relatively isolated
galaxies in the outskirts of the Fornax Cluster, far away from the
brightest ($M_B<-20$) Fornax galaxies (the faint galaxy to the east of
FCC204 in Figure \ref{imafcc204} is FCCB1240, a background
elliptical).

The R- and I-band images were used to extract surface brightness,
position angle and ellipticity profiles. The deviations of the
isophotes from a pure elliptic shape were quantified by expanding the
intensity variation along an isophotal ellipse in a fourth order
Fourier series with coefficients $S_4$, $S_3$, $C_4$ and $C_3$~:
\begin{eqnarray}
I(\theta) &=& I_0 \left[ 1 + C_3 \cos(3\theta)+ C_4
\cos(4\theta)) + S_3\sin(3\theta))+ S_4 \sin(4\theta) \right].
\end{eqnarray}
Here, $I_0$ is the average intensity of the isophote and the angle
$\theta$ is measured from the major axis. If the $4^{\rm th}$
coefficient $C_4$ is positive, the isophotes are ``disky''; otherwise,
they are ``boxy''. If positive, its value is related to the fractional
excess brightness due to the presence of a disk. Else, its value can
be related to the fractional brightness decrement at the ``tips'' of
the boxy isophotes. The photometry of all dEs in our sample will be 
discussed in a forthcoming paper.
\begin{table*}
 \centering \begin{minipage}{140mm} \caption{Basic photometric and
kinematic parameters of FCC204 and FCC288. We used $H_0=75$~km/s/Mpc
(or equivalently a distance $D=18.3$~Mpc).}
\begin{tabular}{@{}llccccccccccc@{}} \hline & type & $m_B^0$\footnote{taken from {\tt NED}} & $m_R^0$\footnote{comparison with {\tt NED}~:~$m_R^0$(FCC288)$=14.33\pm0.9$, $m_R^0$(FCC204)$=13.94\pm0.9$} &
$m_I^0$ & $R_{\rm e}$~(kpc) & $\langle \epsilon \rangle$ & PA
($\degr$) & $n$ & $v_{\rm max}$~(km/s) & $\overline{\sigma}$~(km/s) & $(v_{\rm
max}/\sigma)^*$ \\ \hline FCC204 & dS0(6) & 14.76 & 13.88 & / & 1.03 &
0.60 & 21.9 & 1.29 & 64.6  & 56.0  & 0.75 \\ FCC288 & dS0(7) & 15.10 &14.39 & 13.86 &
0.85 & 0.72 & 4.6 & 1.11 & 60.0 & 39.9  &  0.79 \\ \hline
\end{tabular}
\end{minipage}
\end{table*}

\begin{figure*}[ht]
\vspace*{11cm}
\special{hscale=60 vscale=60 hsize=700 vsize=310
hoffset=20 voffset=330 angle=-90 psfile="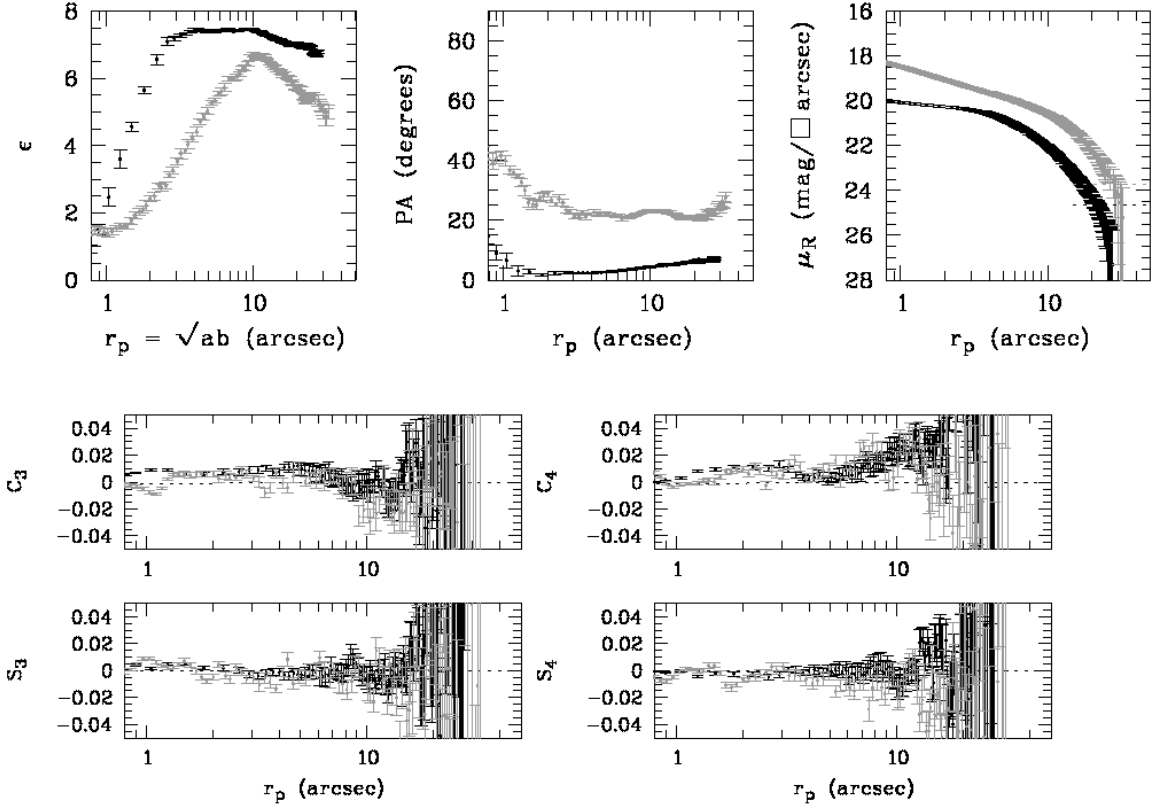"}
\caption{Photometric parameters of FCC288 (black) and FCC204 (grey)
outside the seeing disk. Top row~:~the ellipticity $\epsilon =
10(1-b/a)$, the position angle PA and the R-band surface brightness
$\mu_{\rm R}$ (the dotted lines mark 1\% of the sky intensity). For
clarity, the surface brightness of FCC204 has been offset by $-1$.
Bottom~:~the third and fourth order shape parameters of the isophotes,
$S_4$, $S_3$, $C_4$ and $C_3$. All quantities are plotted as functions
of the geometric mean of the major-axis and minor-axis distances, $a$
and $b$.}
\label{photdS0}
\end{figure*}

The results of this exercise can be found in figure \ref{photdS0}, where
all photometric quantities are plotted as a function of the geometric
mean of the major and minor-axis distances $a$ and $b$. Both galaxies
have positive $C_4$ profiles, i.e. they have disky isophotes which
suggests that they are bulge-disk systems seen practically edge-on. To
check whether this diskiness is indeed caused by the presence of a
real disk, we applied an unsharp masking technique. We smoothed the
R-band image of a galaxy with the {\tt MIDAS} command {\tt
filter/med}\footnote{{\tt ESO-MIDAS} is developed and maintained by
the European Southern Observatory}, which replaces each pixel by the
median of a surrounding box. Like Barazza {\em et al.}, we opted for a
6\arcsec$\times$6{\arcsec} box. This smoothed image is then subtracted
off the original one, highlighting any fine structure. These residual
images are presented in the right panels of Figures \ref{imafcc288}
and \ref{imafcc204}.

\begin{figure}[ht]
\vspace*{7cm}
\special{hscale=40 vscale=40 hsize=700 vsize=650
hoffset=20 voffset=220 angle=-90 psfile="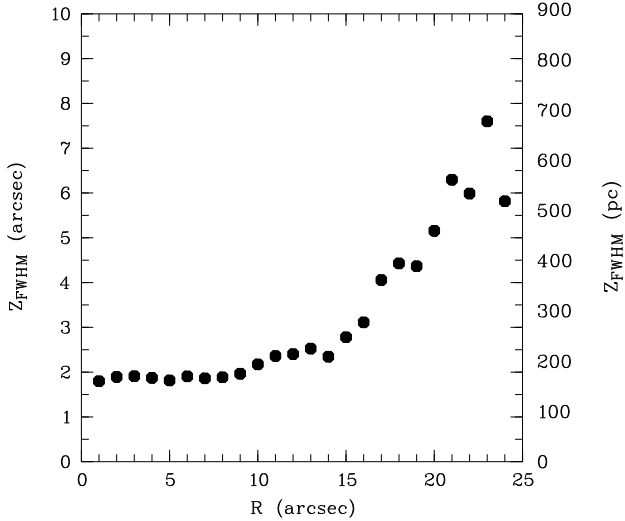"}
\caption{The vertical thickness (${\rm FWHM}_{\rm true}$) of the disk
embedded in FCC288 as a function of radial distance. On the right
side, the corresponding linear distance scale (in parsecs) is
added. \label{thick}}
\end{figure}

The very prominent disk in FCC288 can be traced out to $\pm
23${\arcsec} (2.05~kpc). Clearly visible in Figure \ref{imafcc288} is
the flaring of the disk, i.e. an increase of the scale-height towards
larger radii. The disk is also slightly warped with the north side
tending towards the east. We estimated its thickness at a given radial
distance from the center by fitting a Gaussian to its vertical
profile. The true thickness (in a FWHM sense) can be derived from the
measured ${\rm FWHM}_{\rm obs}$, taking into account the broadening by
the seeing using the relation \[ {\rm FWHM}_{\rm true} = \sqrt{ {\rm
FWHM}_{\rm obs}^2 - {\rm FWHM}_{\rm seeing}^2 }. \] The measured
thickness of the disk remains more or less constant around ${\rm
FWHM}_{\rm obs}=2\arcsec$ (${\rm FWHM}_{\rm true} = 160$~pc) inside
the inner 10{\arcsec}. Beyond that, the disk thickens rapidly reaching
${\rm FWHM}_{\rm obs} \approx 6\arcsec$ (${\rm FWHM}_{\rm true}
\approx 500$~pc) at a radial distance of 20{\arcsec} (Figure
\ref{thick}). In the disk, brightness enhancements can be discerned at
5{\arcsec} and 9{\arcsec} to the north of the nucleus and at
7{\arcsec} and 12{\arcsec} to the south of it (marked by arrows in
Figure \ref{imafcc288}). These could signal the presence of spiral
arms in the disk.

The vertical scale-height of a disk remains constant only if the
vertical velocity dispersion and the surface density drop off in a
delicate balance. In the case of an exponential disk with constant
$M/L$, this is quantified by the relation $h_{\sigma_z} = 2 h_R$
between the scale-length of the vertical velocity dispersion,
$h_{\sigma_z}$, and that of the surface brightness, $h_R$ (van der
Kruit~\&~Searle, \cite{van81}). If this balance is imperfect,
the disk will either flare ($h_{\sigma_z} > 2 h_R$) or taper
($h_{\sigma_z} < 2 h_R$). It is therefore not surprising that flaring
optical galaxy disks are not uncommon (see e.g. Narayan~\&~Jog,
\cite{nar02}; Reshetnikov~\&~Combes, \cite{res98}). Warped optical
disks are discussed by Reshetnikov~\&~Combes (\cite{res98}). These
authors find that about half of all disk galaxies are warped and
suggest that tidal interactions have a large influence in creating or
re-enforcing warped deformations (see also Weliachew {\em et al.},
\cite{wel78}). Bailin~\&~Steinmetz (\cite{bai02}) simulate the
reaction of an optical disk to torques exerted by a misaligned dark
halo and find that a trailing warp develops. If some dEs stem from
harassed disk galaxies, one might expect them to have been stripped of
material at large radii -- including dark matter -- and to have fast
dropping surface-densities and consequently to contain flaring
disks. The tidal forces exerted on a dwarf galaxy during an encounter
can moreover give rise to a warped disk or can induce a misalignment
between disk and halo which also yields a warped appearance.

FCC204 has a much less impressive disk, traceable to about $\pm
20${\arcsec} (1.78~kpc) and about 3.0{\arcsec} thick (${\rm FWHM}_{\rm
true} =255$~pc). Besides a large bulge, two brightness maxima at $\pm
19${\arcsec} from the center are visible (marked by arrows in Figure
\ref{imafcc204}). These might also be signatures of spiral
arms. Another possible interpretation is that we are looking edge-on
onto a bulge+bar system and that the two brightness enhancements, at
symmetric positions with respect to the nucleus, constitute the edges
of the bar or perhaps are even small spiral arms.

\begin{figure*}
\vspace*{7.5cm}
\special{hscale=50 vscale=50 hsize=700 vsize=220
hoffset=20 voffset=280 angle=-90 psfile="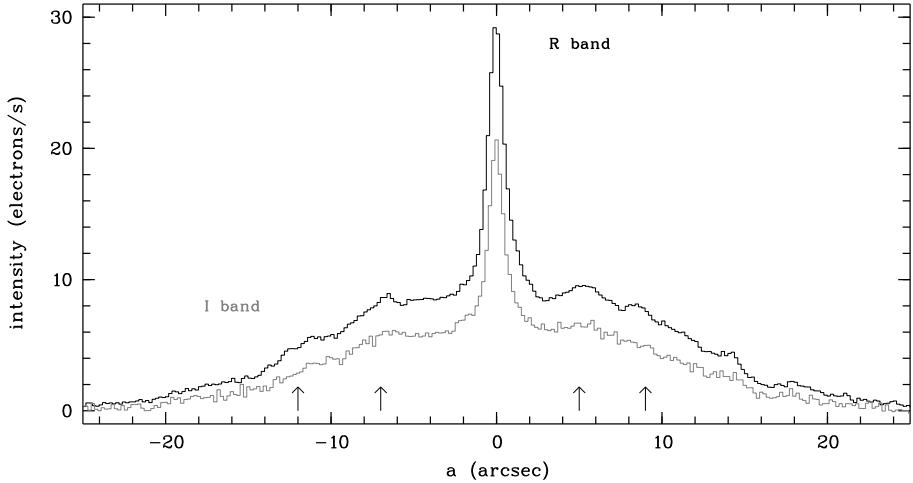"}
\caption{Cut along the major axis of FCC288, black~:~R band, grey~:~I
band. Plotted here is the residual of the unsharp masking. The four
brightness enhancements are marked with arrows. \label{cutfcc288}}
\end{figure*}

\begin{figure*}
\vspace*{8cm}
\special{hscale=50 vscale=50 hsize=700 vsize=220
hoffset=20 voffset=280 angle=-90 psfile="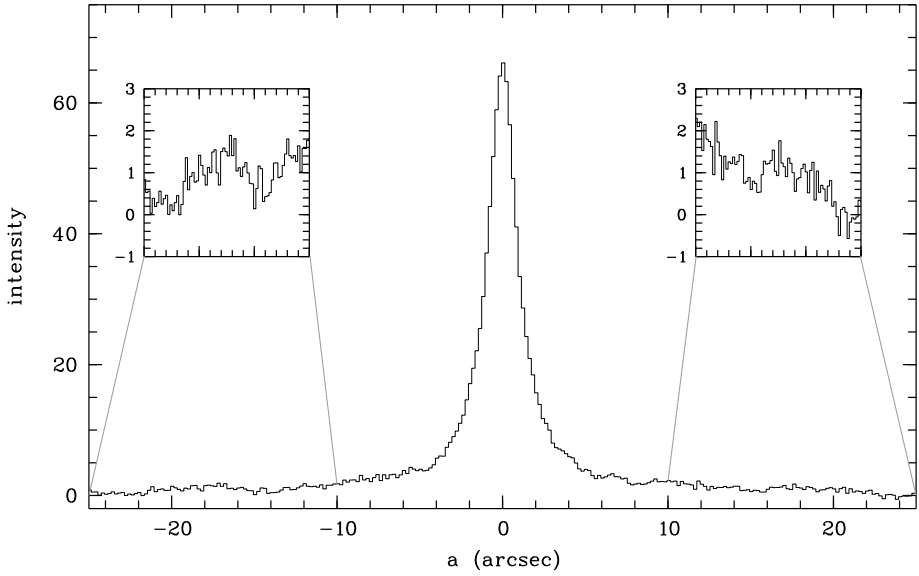"}
\caption{Cut along the major axis of FCC204. Plotted here is the
residual of the unsharp masking. The two insets highlight the
brightness enhancements on symmetric positions about the nucleus.
\label{cutfcc204}}
\end{figure*}

In Figures \ref{cutfcc288} and \ref{cutfcc204}, we plotted the
vertically averaged -- i.e. radial -- R-band luminosity profiles of
the disks in FCC288 and FCC204. Again, one can see the brightness
enhancements in FCC288's disk. Overplotted in Figure \ref{cutfcc288}
is the radial profile of the disk deduced from a 270~sec. I-band
image. Although the seeing in the R and I-band images is the same,
about 0.9{\arcsec}~FWHM, the enhancements seem to be less prominent in
the I-band image, especially the outer ones. Nevertheless, their
showing up also in the I-band image clearly proves that these are not
H$\alpha$ emitting gas clouds which would only be visible in the R
band.


\section{Kinematics} \label{kin}

\begin{figure*}[ht]
\vspace*{10.5cm}
\special{hscale=60 vscale=60 hsize=700 vsize=300
hoffset=20 voffset=350 angle=-90 psfile="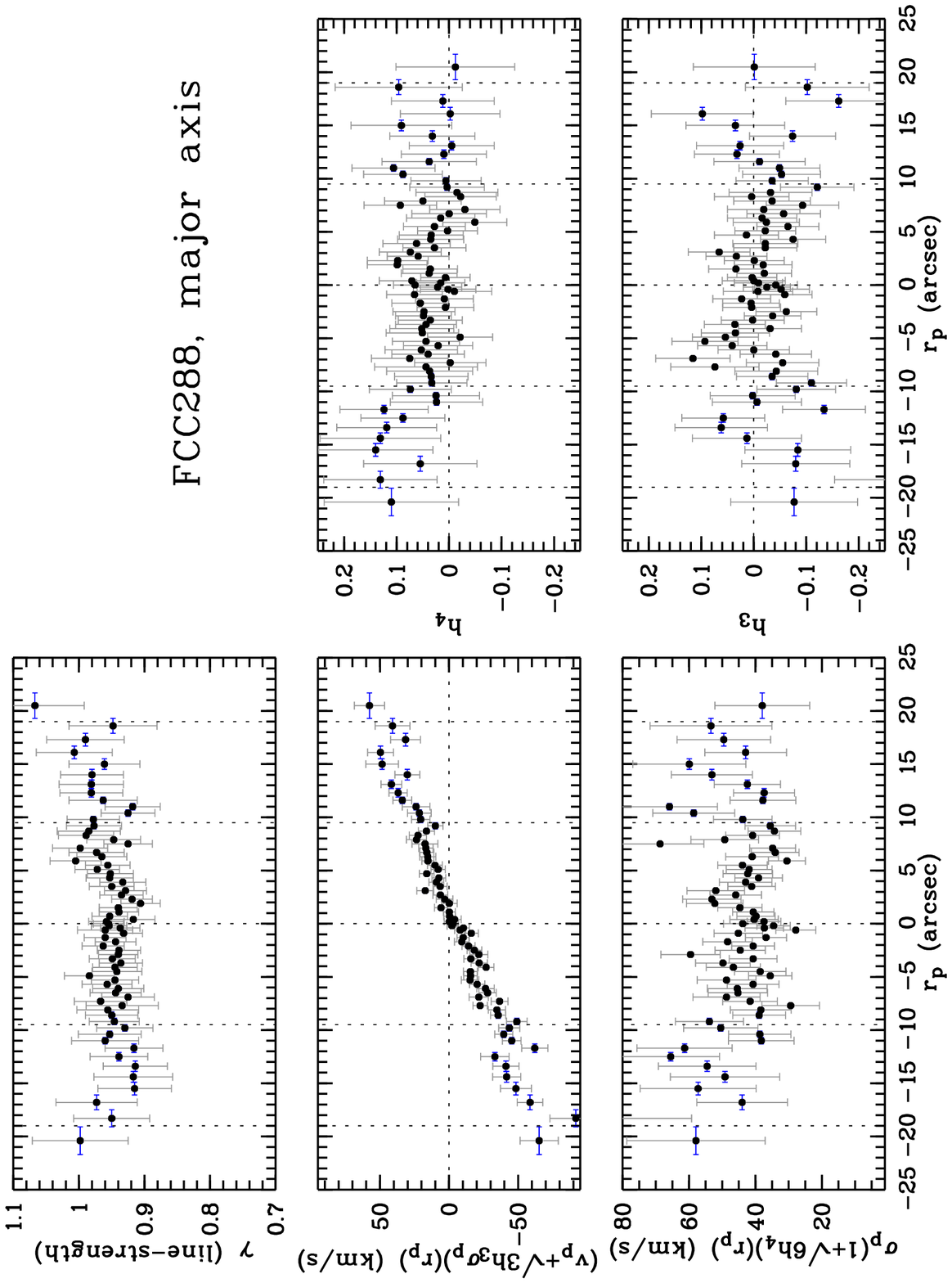"}
\vspace*{10.5cm}
\special{hscale=60 vscale=60 hsize=700 vsize=300
hoffset=20 voffset=350 angle=-90 psfile="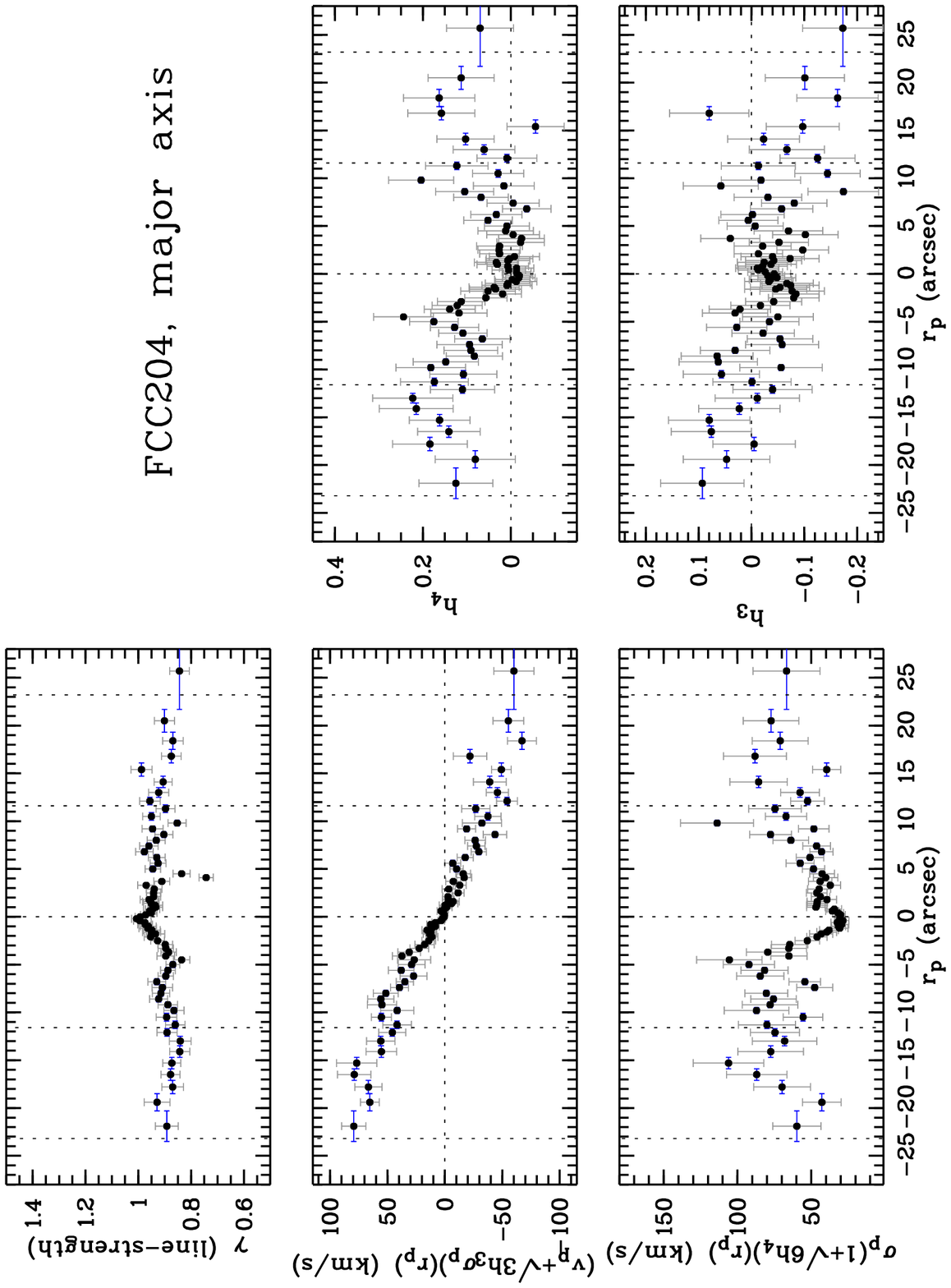"}
\caption{Major axis kinematics of FCC288 (top) and FCC204 (bottom).
For an explanation of these quantities, we refer to the text, section
\ref{kin}. The vertical dotted lines mark $\pm 1$ and $\pm 2$
effective radii. \label{kinfcc288}}
\end{figure*}

Deep long-slit spectra of FCC288 and FCC204 were obtained with {\tt
FORS2} in the wavelength region around the Ca{\sc ii} triplet during
the periods 1-8/11/2000 on Kueyen ({\tt VLT-UT2}) and 11-21/11/2001 on
Yepun ({\tt VLT-UT4}), respectively. We employed the very efficient
holographic grism {\tt GRIS\_1028z$+$29} (throughput $\approx 90$\%
around 8500{\AA}) with a 0.7{\arcsec} slit, resulting in an
instrumental broadening of $\sigma_{\rm instr} = 30$~km/s. The Doppler
broadening of the absorption lines of each row of a galaxy spectrum
was modeled with a line-of-sight velocity distribution (LOSVD) of the
form
\begin{equation}
\phi(v_p) = \frac{\gamma}{\sqrt{2 \pi} \sigma_p} \exp \left( \frac{1}{
2} \left( \frac{v_p-\langle
v_p \rangle}{\sigma_p} \right)^2 \right) \left[ 1 + h_3 H_3(v_p) + h_4 H_4(v_p) \right]
\end{equation}
with $H_3$ and $H_4$ the third and fourth order Hermite polynomials
(van der Marel~\&~Franx, \cite{mar93}). These take into account
asymmetric and symmetric deviations of the LOSVD of a pure Gaussian
profile. A positive $h_4$ yields a LOSVD that is more peaked than a
Gaussian whereas a LOSVD with negative $h_4$ is more flat-topped. A
one-dimensional galaxy spectrum can hence be written as the weighted
sum of a number of broadened template spectra $s_i$ (typically late
G{\sc iii} to late K{\sc iii}) and a few low-order polynomials to take
into account small differences between the continuum of the galaxy
spectrum and that of the sum of broadened stellar spectra~:
\begin{equation}
g(\lambda) = \sum_i c_i \int s_i(\lambda(v_p,\lambda)) \, \phi(v_p)
\, d v_p + {\rm continuum},
\end{equation}
with $\lambda(v_p,\lambda)$ the wavelength that is Doppler-shifted to
$\lambda$ if a star moves at a velocity $v_p$ along the line of
sight. The coefficients $c_i$ are fitted using only the central row of
the galaxy spectrum and are kept constant while fitting to the other
rows. A quadratic programming technique was used to ensure that the
$c_i$ are positive. First, a Gaussian LOSVD is fitted to a 1D galaxy
spectrum. The parameters $\langle v_p \rangle$ and $\sigma_p$ are then
kept fixed to fit the coefficients $\gamma$, $h_3$ and $h_4$ with a
simple least-squares technique. It can be shown that the quantities
\begin{eqnarray}
\langle v_p \rangle' &=& \langle v_p \rangle + \sqrt{3} h_3 \sigma_p \\
\sigma_p' &=&  \sigma_p (1 + \sqrt{6} h_4)
\end{eqnarray}
are better approximations of the mean line-of-sight velocity and the
velocity dispersion of the LOSVD than the Gaussian estimates. These
kinematic parameters are plotted in Figure \ref{kinfcc288} as a
function of position along the major axes of these galaxies. The
kinematics of all dEs in our sample will be discussed in a forthcoming
paper.

Clearly, both galaxies are very rapid rotators compared to most dwarfs
of comparable luminosity. The luminosity-weighted average velocity
dispersion of FCC288 is $\sigma_{\rm aver} = 39.9 \pm 3.9$~km/s, that
of FCC204 is $\sigma_{\rm aver} = 47.0 \pm 5.9$~km/s. The maximum
rotation velocities are estimated at $v_{\rm max} = 60.0 \pm 4.0$ and
$v_{\rm max} = 62.4 \pm 3.6$ for FCC288 and FCC204, respectively. The
anisotropy parameter $(v_{\rm max}/\sigma_{\rm aver})^*$ is defined as
\begin{equation}
(v_{\rm max}/\sigma_{\rm aver})^*= (v_{\rm max}/\sigma_{\rm
aver})_{\rm obs} / (v_{\rm max}/\sigma_{\rm aver})_{\rm theo}
\end{equation}
with $(v_{\rm max}/\sigma_{\rm aver})_{\rm obs}$ the observed ratio
and $(v_{\rm max}/\sigma_{\rm aver})_{\rm theo}$ the one expected if
the galaxy would be an isotropic, oblate galaxy flattened by
rotation. Hence, if a galaxy is flattened purely by rotation, one
would expect $(v_{\rm max}/\sigma_{\rm aver})^*=1$. For FCC288, we
find $(v_{\rm max}/\sigma_{\rm aver})_{\rm obs} = 1.50 \pm 0.25$ and
FCC204~:~$(v_{\rm max}/\sigma_{\rm aver})_{\rm obs} = 1.33 \pm
0.24$. Using equation (4-95) from Binney~\&~Tremaine (\cite{BT87}) to
calculate $(v_{\rm max}/\sigma_{\rm aver})_{\rm theo}$, one then finds
that $(v_{\rm max}/\sigma_{\rm aver})^*($FCC288$)=0.79 \pm 0.13$ and
$(v_{\rm max}/\sigma_{\rm aver})^*($FCC204$)=0.86 \pm 0.16$. Both
galaxies have $(v_{\rm max}/\sigma_{\rm aver})^*<1$. This cannot be
due to inclination~:~they are seen practically edge-on. It is most
likely due to the fact that the kinematics are influenced by the stars
that make up the less flattened (i.e. slower rotating, more
anisotropic) body of the galaxy. However, these values are still very
high, compared to other dEs. Most dEs observed so far show no or very
little rotation (e.g. Geha {\em et al.}  \cite{geh02}).

Whereas $\gamma$, which controls the depth of the absorption lines,
$h_4$ and the velocity dispersion are more or less constant along the
major axis of FCC288, they vary significantly inside the inner few
arcseconds in the case of FCC204. This is most likely due to presence
of the strong bulge in FCC204. The LOSVDs of both disks are quite
peaked with $h_4 \approx 0.05$ in the case of FCC288 and $h_4 \approx
0.2$ in that of FCC204. Inside the bulge of FCC204, the LOSVDs have a
Gaussian shape. In the disk, the LOSVDs are more peaked and
skewed. Beyond about $\pm 20${\arcsec}, i.e. at the outer edges of the
putative bar, the velocity dispersion starts to decline and the
rotation velocity levels off. The very fast rotation and the behavior
of the various kinematic parameters strengthen our interpretation of
the photometrically identified features~:~both galaxies are seen
practically edge-on and contain fast-rotating disk structures. The
edges of what we interpret to be a bar in FCC204 correspond with
changes in the velocity dispersion and the rotation velocity.

\section{Discussion and conclusions} \label{con}

We have presented photometric and kinematic evidence for embedded
disks in FCC288 and FCC204, two bright dwarf galaxies in the Fornax
Cluster. They were classified as dS0(7) and dS0(6) respectively on the
basis of their disky isophotes. These were the only dwarfs that showed
any sub-structure out of a sample of 22 dEs (both nucleated and
non-nucleated and with a wide variety of ellipticities) from the
Fornax Cluster and the NGC5044 and NGC5898 Groups. FCC288 has a very
prominent disk, extending to about 2~kpc at either side of the galaxy
center, but has only a small bulge, like a late-type disk
galaxy. Inside the disk, brightness enhancements can be discerned,
which we interpret as traces of spiral arms. The disk flares at larger
radii and is slightly warped. FCC204 on the contrary has a large bulge
but a much less impressive disk, somewhat like an early-type disk
galaxy, extending to about 1.8~kpc, ending in two bright spots on
symmetric positions at either side of the bulge. A possible
interpretation is that FCC204 contains a bar that ends in two bright
knots where small spiral arms start. The kinematic data, particularly
the very fast rotation and the correlations between the photometric
and the kinematic features, corroborate these interpretations. All
observed features can be understood in the context of the
galaxy-harassment scenario. If some dEs stem from harassed disk
galaxies, they might still contain relics (such as embedded disks) of
their former state. The observed flaring and warping of the disk in
FCC288 might be consequences of the interactions during the
harassment.

Since at the time of writing no dEs with embedded disks were known in
other clusters, Barazza {\em et al.} (\cite{bar02}) noted that Virgo
might be somewhat special. It is a very massive and extended cluster
(Madore~\&~Freedman, \cite{mad98}) and galaxies are falling into it at
a constant rate (Binggeli {\em et al.}, \cite{bin87}). The dEs with
embedded disks could then very well be relatively new members of the
cluster, distinct from the preexisting dEs, that are currently in the
process of being transformed from late-type into early-type
dwarfs. Fornax however is less massive and more spatially
concentrated than Virgo. Unlike Virgo, it has no continuous infall of
galaxies. Therefore, dEs with properties reminiscent of late-type
galaxies are most likely formed by harassment of disk galaxies
from within the cluster itself. The harrassment timescale should then
be sufficiently long that they have retained evidence of their
original late-type nature.

\begin{acknowledgements}
This research has made use of the NASA/IPAC Extragalactic Database
(NED) which is operated by the Jet Propulsion Laboratory, California
Institute of Technology, under contract with the National Aeronautics
and Space Administration. WWZ acknowledges the support of the Austrian
Science Fund (project P14783) and of the Bundesministerium f\"ur
Bildung, Wissenschaft und Kultur.
\end{acknowledgements}

\end{document}